\documentclass[twocolumn]{aastex631}
\usepackage{amsmath}

\begin{document}

\title{Low-latitude magnetic flux emergence on rapidly rotating solar-type stars}

\correspondingauthor{Emre I\c{s}{\i}k}
\email{[isik,solanki,cameron,shapiroa]@mps.mpg.de}

\author[0000-0001-6163-0653]{Emre I\c{s}{\i}k}
\affiliation{Max-Planck-Institut f\"ur Sonnensystemforschung \\
              Justus-von-Liebig-Weg 3, 37077 G\"ottingen, Germany}

\author[0000-0002-3418-8449]{Sami K. Solanki}
\affiliation{Max-Planck-Institut f\"ur Sonnensystemforschung \\
              Justus-von-Liebig-Weg 3, 37077 G\"ottingen, Germany}

\author[0000-0001-9474-8447]{Robert H. Cameron}
\affiliation{Max-Planck-Institut f\"ur Sonnensystemforschung \\
              Justus-von-Liebig-Weg 3, 37077 G\"ottingen, Germany}

\author[0000-0002-8842-5403]{Alexander I. Shapiro}
\affiliation{Max-Planck-Institut f\"ur Sonnensystemforschung \\
              Justus-von-Liebig-Weg 3, 37077 G\"ottingen, Germany}



\begin{abstract}
Besides a dense coverage of their high latitudes by starspots, rapidly rotating cool stars also display low-latitude spots in Doppler images, although generally with a lower coverage. 
In contrast, flux emergence models of fast-rotating stars predict strong poleward deflection of radially rising magnetic flux as the Coriolis effect dominates over buoyancy, leaving a spot-free band around the equator. 
To resolve this discrepancy, we consider a flux tube near the base of the convection zone in a solar-type star rotating eight times faster than the Sun, assuming field intensification by weak-tube explosions. 
For the intensification to continue into to the buoyancy-dominated regime, the upper convection zone must have a significantly steeper temperature gradient than in the Sun, by a factor that is comparable with that found in 3D simulations of rotating convection. 
Within the hypothesis that stellar active regions stem from the base of the convection zone, flux emergence between 1-20 degree latitudes requires highly supercritical field strengths of up to 500 kG in rapidly rotating stars. These field strengths require explosions of 100-kG tubes within the convection zone, compatible with reasonable values of the superadiabatic temperature gradient associated with the more rapid rotation.
\end{abstract}




\section{Introduction} \label{sec:intro}

Photospheres of cool stars with rotation periods of up to a few days 
are dominated by high-latitude or polar darkenings associated with 
magnetic flux concentrations \citep{strass09r}. 
Unseen on the Sun, this is interpreted as the result of a combination of 
two effects that, in principle, can work in parallel. Firstly, at a sufficient 
rotation rate, the Coriolis acceleration experienced by rising flux tubes 
can overcome the radially outward directed acceleration caused by buoyancy 
\citep{schuessler92}. Secondly, at a sufficient emergence frequency, 
trailing-polarity flux from tilted bipolar magnetic regions emerging at any 
latitude diffuse and accumulate around polar regions. For very high levels of magnetic flux emergence, which lead to high activity levels, this accumulated flux is expected to form dark polar features \citep{schrijver01} with enhanced lifetimes 
\citep{isik07}, or mixed-polarity polar spots if the meridional flow is much faster than on the Sun \citep{holzwarth06}. 

In spite of the apparent success of models 
reproducing starspots at high latitudes and poles of rapid rotators, numerical simulations of 
flux emergence all predict a `zone of avoidance' between about $\pm 20^\circ$ 
latitudes, where no flux emergence occurs \citep{msch96,granzer00,isik11,isik18}. 
However, Doppler and Zeeman-Doppler imaging (hereafter DI and ZDI, respectively) 
reconstructions of rapidly rotating cool stars show occasional low-latitude 
spots \citep[e.g.,][]{rice+strass01,jeffers+02,perugini21,willamo22}. 
It is known that the resolving power of Doppler imaging is rather low in 
latitude. In addition, there is a possible information leak from mid-latitude 
activity on the less visible rotational hemisphere, depending on the 
axial inclination \citep{senavci21}. 
Nevertheless, low-latitude magnetic features are seen in most (Z)DI 
reconstructions of fast rotators, even though the latitude profile of 
the longitudinally averaged spot occupancy sharply drops equatorward of 
$\sim 20^\circ$ latitude in most cases. 

\begin{figure}
    \centering
    \includegraphics[width=.9\columnwidth]{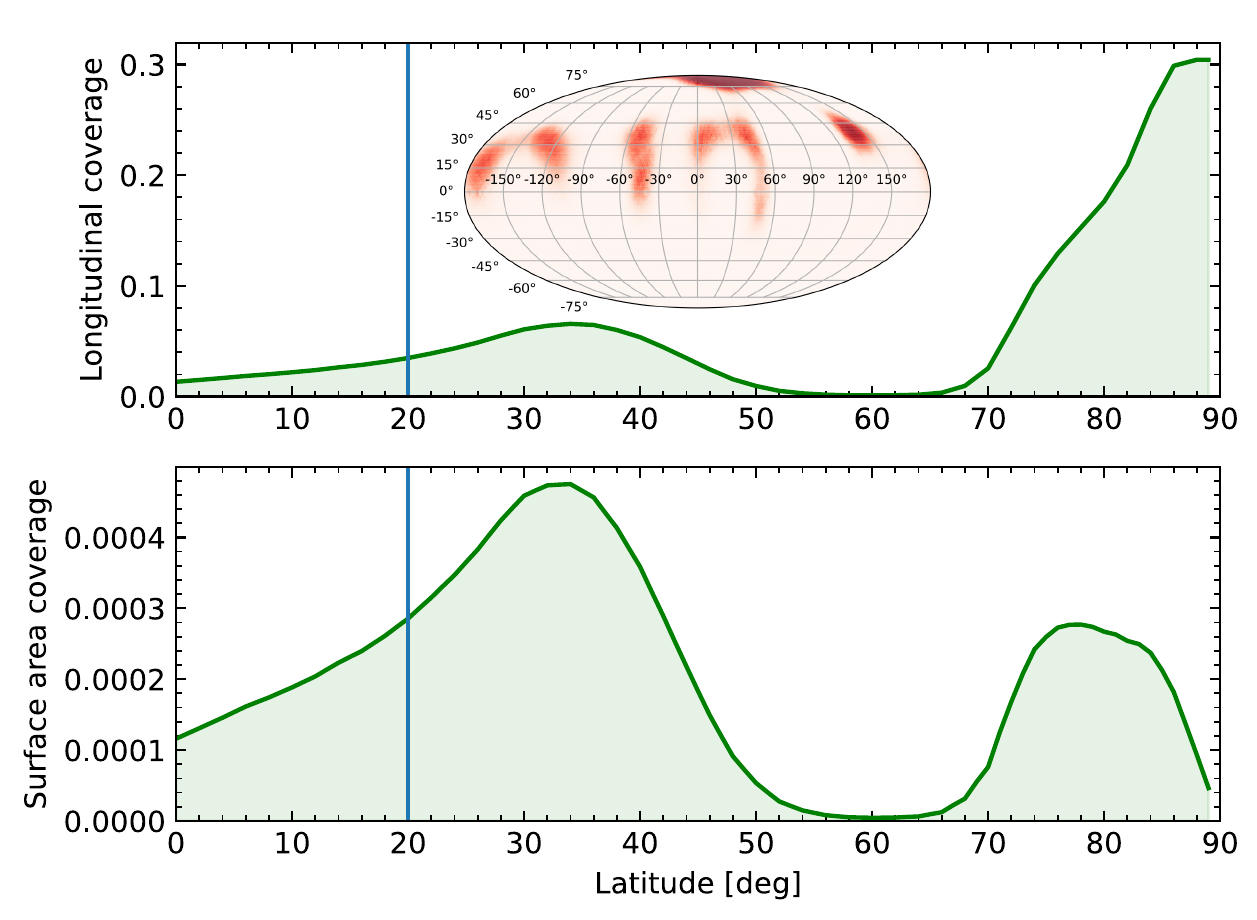}
    \caption{The latitudinal profile of longitude-averaged latitude occurrence (top panel) and the surface fraction of spots per degree of latitude on EK Dra in July 2015. The inset shows the surface reconstruction 
    of spot filling factor with Doppler imaging. Note that the stellar inclination is $63^\circ$, so that no spots are visible on one rotational hemisphere. The image data was taken from \citet{senavci21}.}
    \label{fig:EKDraprofile}
\end{figure}

An example latitude distribution is shown in Fig.~\ref{fig:EKDraprofile}, using data from \citet{senavci21}. The latitudinal profile of relative spot occupancy is averaged over longitude, using the Doppler image reconstructed from a 15-night observing campaign.
The top panel shows the latitude occurrence of spotted pixels on the map. 
The bottom panel shows the surface area per degree of latitude, which is covered by spot pixels, normalised to the surface area of the star. 
While there is a near-polar spot region spanning 70-90$^\circ$ latitudes, 
the mid-latitude spots peak 
around $33^\circ$, with an elevated tail towards the equator. Spots below 
$20^\circ$-latitude are visible in the surface reconstruction (inset 
image). Note that low-latitude features with cross-equatorial extensions are likely artefacts of mid-latitude activity on the less visible rotational hemisphere \citep{senavci21}. Nonetheless, a number of the low-latitude spots are likely real. A theoretical picture of how such low-latitude features form amid rapid rotation is not yet available. 

On the Sun, low-latitude flux emergence has important effects on the cross-equatorial transport of active-region dipole moments and hence for the build-up and long-term fluctuations of the global axial dipole moment \citep{cameron13}. 
Active regions emerging near the equator of a rapidly rotating active star can also drive a solar-like dipolar dynamo mode. 

The photospheric magnetic field of the Sun is known to be far from uniform. It consists of a hierarchy of flux tubes that are organised 
into spots, pores, and smaller-scale concentrations \citep{solanki06}. 
Sunspot groups emerge as a consequence of the rise of bundles of magnetic 
flux from the solar convection zone (hereafter CZ) \citep{weber23}.
One model that reproduces several important characteristics of flux emergence on scales of sunspot groups 
makes use of numerical solutions of thin-flux-tube equations of ideal MHD. The model considers a toroidal ring that carries a 
similar magnetic flux as large solar active regions located near the 
base of the CZ. There the stratification is expected to be sufficiently sub-adiabatic to allow for mechanical equilibria in the presence of an internal 
flow along the tube axis. The tube has a sufficiently small cross section in comparison to the local pressure scale height (over which the thermodynamic quantities 
undergo significant changes). Linear stability analysis \citep{ferriz95} helped develop numerical models of flux-tube dynamics spanning the CZ \citep{cale95}, which led to a series of 
results that are consistent with the observed trends of morphological 
and geometric properties of sunspot groups. 
For higher rotation rates and deeper convection zones (i.e., for 
lower Rossby numbers), the dynamical flux-tube simulations turned out to be 
consistent with the observation that mean latitudes of starspots are higher 
than on the Sun. 
However, when the initial flux-tube field strengths were taken following the approach in solar simulations (ie., slightly above the onset of buoyancy instability that carries a flux tube to the surface), the resulting emergence patterns failed to reproduce low-latitude spots observed on fast rotators. 

One poorly understood aspect is the depth of the original flux ring that gives 
rise to the eruption of a flux loop. To date, there has not been any conclusive result for the origin of the emerging flux loops on the Sun. Linear stability constraints hint towards the stably 
stratified lower part of the CZ as the seat of the flux tubes producing active-regions \citep{fan21}. However, 3D simulations of 
magnetoconvection in a convective shell rotating three times faster than the Sun led to the formation of toroidal wreaths amid turbulent convective flows \citep{nelson11,nelson14}. 
This result is consistent with a distributed dynamo, in which active-region 
producing flux tubes stem from a wide range of depths in the CZ. On the one hand, these simulations include much detail and insight into the physics of flux storage and emergence. On the other hand, the Reynolds number reached in such simulations is still far below that present within a stellar CZ. Furthermore, a subadiabatic lower convection zone (i.e., connecting to a radiative zone), which would facilitate flux storage also near the base, was not included in these models. However, it is relatively easier to provide statistics 
of flux emergence patterns with hundreds of thin flux tube simulations having different initial conditions 
(latitudes, field strengths, depths), considering the main bulk forces acting on any rising flux concentration in 3D with a super-equipartition field. This approach is currently useful for modelling stellar magnetic activity with increasing rotation rate, until more realistic 3D magnetoconvection simulations for different rotation rates will become available. 

We therefore study the intensification and breakup 
of thin toroidal flux tubes at the base of the CZ, at a 
rotation rate of eight times the solar rate ($P_{\rm rot}=3.125$~d), 
firstly by assuming a solar-like internal stratification based on 
non-local mixing-length theory, followed by a modified stratification 
adapted for the case of rapid rotation. 
Because the dynamo-generated magnetic flux in rapidly rotating stars is expected to be higher than on the Sun (as suggested by the empirical rotation-activity relation), it is conceivable that the flux density at the source depth can also reach higher levels of supercriticality than the near-critical levels for buoyancy instability. Can supercritical flux tubes form in the first place, and how? 

We begin by considering a flux tube that is about 30 Mm above the base of the CZ, 
where the stratification turns from subadiabatic (entropy increasing outward) 
at the lower CZ to superadiabatic above. Here, the critical field strength for magnetic 
buoyancy instability is on the order of equipartition fields ($10^4$~G). 
When such a tube develops a loop that rises isentropically, the internal gas 
pressure at its apex will fall slower than the ambient pressure, until 
when the lateral pressure balance 
can be satisfied only if $B\rightarrow 0$, so the tube explodes 
\citep{moreno95,rempel01,hotta12}. Following drainage of material out of 
the exploded part, the cross-sectional area of such a tube will shrink along 
its sunken parts in the overshoot region, 
with a corresponding increase in magnetic field strength, 
on a time scale $t_f$ which is short compared to the local timescale of the buoyant 
instability $t_g$. As the field becomes stronger, $t_g$ changes faster than 
$t_f$, and at some critical field strength $t_f=t_g$ holds and the 
magnetic buoyancy instability becomes the dominant process, in the sense that the flux escapes faster than it is produced. 
We note that $t_f$ depends on the superadiabaticity (the difference of the logarithmic 
temperature gradient from the corresponding adiabatic gradient) in the bulk of the convection zone 
and this is thought to be dependent on the rotation rate (Sect.~\ref{ssec:growthtimes}). 
We therefore argue that 
the initial conditions for the flux tube simulations should depend on the 
stellar rotation rate. With plausible values of the field strength and the 
superadiabaticity, we find that flux loops producing starspot groups can 
($i$) originate from the base of the CZ and ($ii$) lead to low-latitude emergence even for 
rapidly rotating stars.

In this study, we investigate the physical ranges of the field strength that can be reached, using the flux tube intensification mechanism outlined above (Sect.~\ref{ssec:growthtimes}). Taking the resulting constraints into account, we place toroidal flux tubes at
different depths in the overshoot layer below the convection zone of a star rotating at eight times the solar rate at the constrained range of field strengths. We study the effects of varying the initial depth and field strength in Sect.~\ref{ssec:dynamics}. Following that, we investigate the effect of the field strength for various internal differential rotation profiles and amplitudes in Sect.~\ref{ssec:dr}. We provide mapped probability distributions of latitude for the emerging flux, in Sect.~\ref{ssec:patterns}. Discussing implications of our model in Sect.~\ref{sec:discuss}, we summarise our conclusions in Sect.~\ref{sec:conc}.

\section{Field intensification in a fast rotator}
\label{sec:model}

\subsection{The flux-tube model}
\label{ssec:tft}

We set up a toroidal flux tube with cross-sectional radius of $R_{\rm t}$, located in a 
1D non-local mixing-length model of the solar convection zone, using 
the thin-flux-tube approach \citep[see][for details]{isik15}. 
Up to about $0.77R_\odot$, 
the stratification is subadiabatic (ie, locally stable to convective 
instability), but convective heat flux remains upward, driven by 
non-locally determined convective heat flux \citep{skastix91}. 
At about $0.73R_\odot$, the convective 
heat flux changes sign (defined as the base of the CZ, at a radial location $r=r_b$), where the 
model allows for an overshoot region that extends the base of the CZ to about 10~Mm depth into the 
radiative interior.
In the standard case, the stellar rotation is assumed to have the 
same differential rotation magnitude ($\Delta\Omega$) as in the Sun, 
but an equatorial rotation rate 8 times faster 
($P_{\rm rot}=3.125$~d; we discuss the effect of differential rotation in Sect.~\ref{ssec:dr}).
Following \citet{ferriz95}, we apply the linear stability analysis of a 
thin flux tube with a field strength $B_0$ in mechanical equilibrium, placed at a given initial depth $d_0=r_b-r$ and latitude $\lambda_0$. This gives the critical field 
strength $B_{\rm cr}(d_0,\lambda_0)$, above which `supercritical' 
flux tubes become unstable to first-order perturbations. In addition, 
the characteristic (e-folding) growth time $t_g$ of the 
instability is calculated for $B_0>B_{\rm cr}(\lambda_0)$, i.e., for supercritical field strengths at a given latitude.

We let the initial field strengths be determined by contours of constant $t_g$ on the 
$(B_0,\lambda_0)$ plane at a given $d_0$. Taking the initial condition to 
be mechanical equilibrium, we follow the nonlinear 
dynamics of flux tubes in 3D, using 
the Lagrangian code of \citet{cale95}, to track their peaks as they 
rise to the surface. The code solves the dynamics of a flux 
ring in ideal MHD, under the effects of body forces (buoyancy, 
Lorentz, and hydrodynamic drag) and rotational 
forces, acting on $10^3$ individual mass elements of the tube in their 
co-moving frame, from the overshoot region up to the subsurface 
($\sim 0.98-0.99R_\star$), where the cross-sectional radius of the 
tube becomes of the order of the pressure scale height, making the 
thin flux tube approximation inapplicable.
The simulations also take into account the differential 
rotation of the stellar convection zone (see Sect.~\ref{ssec:dr} for 
more details on the internal rotation profiles considered).

\subsection{Flux tube explosions in fast rotators}
\label{ssec:growthtimes}

Prior to numerical simulations leading to low-latitude emergence, 
we estimate constraints on $B_0$ in a fast-rotating 
solar-type star, by considering possible effects of flux 
intensification and the convection-zone stratification. 
We assume that flux tubes are formed on a timescale $t_f$
out of a mean toroidal field, which is wound up from a poloidal field 
by differential rotation throughout the convection zone. 
We put a conservative requirement that tubes must reach such field 
strengths sufficiently rapidly at a timescale $t_f\lesssim t_{\rm g}$ 
before the magnetic buoyancy instability lifts them off the 
overshoot region into the convection zone. This is a conservative 
constraint in the sense that the actual time from the initial 
perturbation to the escape of a substantial amount of flux from the 
overshoot region is 
longer than $t_g$, as will be shown in Sect.~\ref{ssec:dynamics}.

To evaluate $t_f$, we consider a mechanism to form strong flux tubes 
at the base of the CZ, suggested by \citet{rempel01}. It
exploits the potential energy available in the stratification of the 
CZ. 
Consider a tube with $B_0$ on the order of the equipartition value
($10^4$~G) and with sufficiently low flux $\Phi_0$, so that drag-braking during 
the rise would help hydrostatic equilibrium to be sustained 
along the tube. When the tube rises isentropically, 
its apex reaches a critical height in the CZ, 
where the internal gas pressure becomes equal to the external 
hydrostatic pressure, leading to $B\rightarrow 0$ and thus 
an explosion of that part of the tube \citep{moreno95}. In the deeply 
buried horizontal parts, however, the field is amplified on the Alfv\'enic 
time scale, owing to draining of plasma out of the exploded part. The relevant
time scale estimated by \citet{rempel01} for the base of the solar 
CZ reads 
\begin{equation}
t_f\simeq (\beta_0 / C)^{1/2},
\label{eq:rempel}
\end{equation}
where $\beta_0$ 
is the plasma beta (associated with the field strength of the original 
flux ring) and $C=5\times 10^7~{\rm yr}^{-2}$ is a constant, determined 
by the sound speed at the base of the CZ.

\begin{figure}
    \centering
    \includegraphics[width=\linewidth]{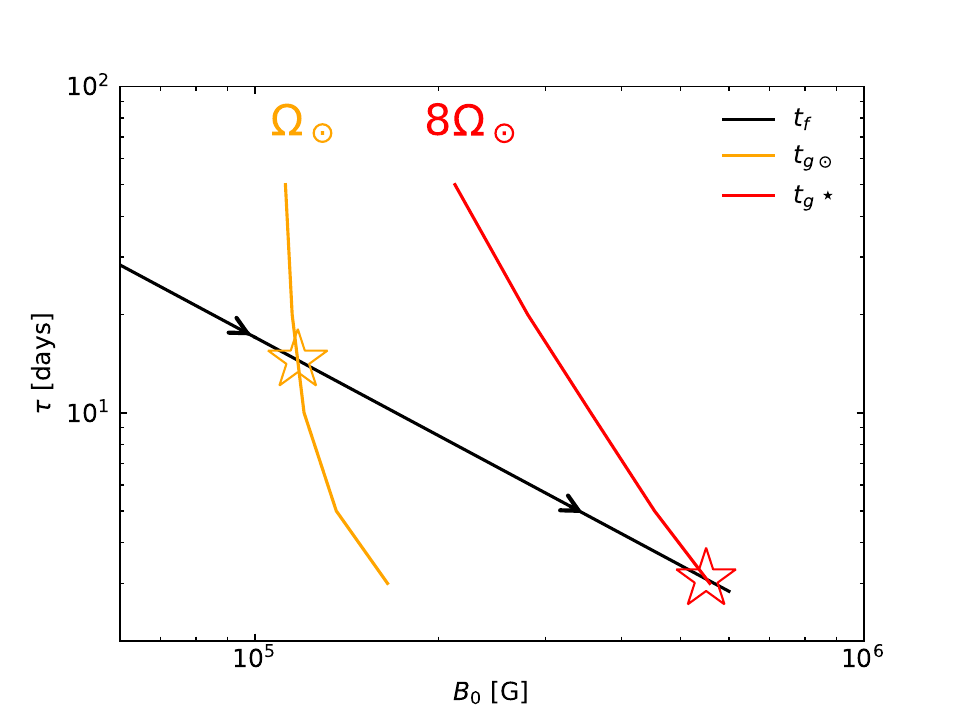}
    \caption{The intensification time $t_f$ as a function of the initial field strength inside a flux tube according to Eq.~(\ref{eq:rempel}), compared to the linear instability growth time $t_g$ of magnetic flux tubes initially located at 5~Mm below the
     base of the convection zone at a latitude of $1^\circ$, in a star rotating at $\Omega_\odot$ (orange) and $8\Omega_\odot$ (red and other colours). As $B_0$ increases along the $t_f$ curve with flux-tube explosions, the tubes are expected to become unstable and rise to the surface around field strengths marked by star symbols.}
    \label{fig:time_rempel}
\end{figure}

The scaling in Eq.~(\ref{eq:rempel}) for the amplification time $\tau:=t_f$ as a 
function of $B_0$ is indicated in Fig.~\ref{fig:time_rempel} 
by the black curve. In the solar convection zone, the explosion mechanism 
can work for magnetic fluxes 
on the order of $10^{17}$~Mx and for fields of up to a few times $10^4$~G. 
The reason is that the explosion height $r_e$ increases with the initial field 
strength of the tube, approximately following 
\begin{equation}
    r_e = r_0 + H_0\left(\frac{\beta_0\delta_s}{2}\right)^{-1/2},
    \label{eq:explosion}
\end{equation}
where $H_0$ is the pressure scale height at the initial location ($r=r_0$) and 
$\delta_s>0$ is the mean superadiabaticity ($\delta(r):=\nabla(r)-\nabla_{\rm ad}$) 
in the unstably stratified part of the convection zone, which is assumed to be constant 
in this basic estimation \citep{moreno95}. Concerning field intensification, an 
explosion at $r_e$ given by Eq.~(\ref{eq:explosion}) can only make sense if 
$r_e < R_\odot$. Since the right-hand side is proportional to $B_0\delta_s^{-1/2}$, 
the main factors determining the explosion radius are $B_0$ and $\delta_s$. 
Assuming $\delta(r)$ to be the solar profile in our adopted stratification model, 
$B_0$ would then be strictly limited to $B_0\lesssim 10^5$~G, 
to satisfy $r_e<R_{\odot}$. For stronger initial fields, the internal and external gas 
pressures do not match within the convection zone, so the tube reaches the surface, 
without leading to major intensification at its roots.
For flux tubes starting at $d_0=5$~Mm and $\lambda_0=1^\circ$ in the Sun, 
the buoyancy instability sets in at $B_{\rm cr}\sim 10^5$~G. Parker-unstable tubes that would carry 
sunspot-group fluxes have characteristic growth times $\tau:=t_g$, following the orange 
curve in Fig.~\ref{fig:time_rempel}. 

Increasing the rotation rate to $8\Omega_\odot$, tubes with $B_0\simeq 10^5$~G
become stable to perturbations at $d=5$~Mm 
\citep[except for friction-induced instability acting on $t\gg t_f$;][]{holzwarth08}. 
$B_{\rm cr}$ and the corresponding 
$B_0$ range of unstable tubes with the same growth times are thus shifted to higher 
values, owing to the stabilising effect of rotation. The resulting range of $B_0$ for emerging 
flux tubes in $8\Omega_\odot$ is thus well beyond the range permitted by the 
$r_e<R_\star$ criterion of the explosion mechanism. Though the radius estimate 
in Eq.~(\ref{eq:explosion}) is based on simple approximations, we confirmed it by 
a set of numerical simulations, where explosions took place inside the CZ only if 
$B_0<10^5$ (and for $\Phi_0\lesssim 10^{18}$~Mx). 

We assumed in the above discussion that $\delta_s$ in Eq.~(\ref{eq:explosion}) 
does not change with $\Omega_\star$. 
However, stars rotating faster than the Sun are expected to have more 
superadiabatic outer CZs, i.e., they have less efficient convection than in the 
Sun \citep{kapyla05,barker14,augustson19}. 
In simulations by \citet{kapyla05}, the effects of rotation and convection are 
quantified by the Coriolis number, ${\rm Co}=2\Omega\tau_c$, where $\tau_c$ is the 
convective turnover time. For instance, a ten-fold change from Co=1 
to Co=10 (equivalent to a ten-fold increase in $\Omega$ at a constant $\tau_c$) 
raises $\delta_s$ by a factor of 3 in the bulk of the upper CZ (see their 
Fig.~5). Eq.~(\ref{eq:explosion}) would then indicate a decrease in 
the explosion radius $r_e$ with increasing $\Omega_\star$. We tested this effect, 
by modifying the surface temperature to be by 50-100 K cooler with a 
smooth transition from the interior, and 
calculated the resulting changes in pressure, density, and $\delta(r)$ 
\citep{rempel03,isik15}, 
resulting in an increase of $\delta_s$ by a factor of 3-10 in the upper CZ. 
Simulations using that modified stratification have led flux tubes with 
$B_0=7-10\times 10^4$~G to explode between $0.95-0.97 R_\star$. 

Following an explosion at the tube apex, the field strength 
in the overshoot region can be amplified (on timescale $t_f$) by a factor of 3-5 
in the 2D flux-sheet modelled by \citet{rempel01}, depending on the initial field 
strength. In 3D simulations, \citet{hotta12} found a factor of 2-3. 
When a $10^5$-G tube explodes at its apex in a fast rotator, 
its roots can thus be amplified towards the $B_0$ values required for 
the buoyancy instability (Fig.~\ref{fig:time_rempel}).

Although uncertainties remain for the internal stratification \citep{hotta22}, we 
consider the possibility that an enhanced superadiabaticity in the mid- to upper CZ of 
rapidly rotating CZs can make the field-amplification curve 
in Fig.~\ref{fig:time_rempel} a viable path 
towards 500-kG fields near the base of the CZ of a star rotating at $8\Omega_\odot$, 
if the tube is amplified mainly by the non-local explosion mechanism. 
Downward pumping of magnetic flux driven by a turbulent diffusivity gradient 
is also a possible pathway contributing to formation of intense magnetic flux tubes 
\citep[e.g.,][]{barker12}.

Consider a flux tube located at $d_0=5$~Mm in a solar-type star with 
$8\Omega_\odot$ being intensified up to a few times $10^5$~G by flux-tube explosions. 
If $B_0\lesssim 500$~kG, the tube can develop buoyancy instability at the timescale 
$t_g(B_0)>t_f(B_0)$ 
(the red curve in Fig.~\ref{fig:time_rempel}), when explosions are no longer possible. 
The condition $t_g\sim  t_f$ (red star symbol in Fig.~\ref{fig:time_rempel}) 
puts another upper limit on the field strength that can be reached 
by flux-tube explosions on this fast rotator, because for $t_g\lesssim t_f$, the 
flux will be rapidly lost from the layer. 
The magnitude of this limit is comparable to the one implied by the explosion criterion 
$r_e<R_\star$ with $\delta_s(\Omega_\star)$ discussed above. 

This comparison simply shows 
the expected mean behaviour and thus provides a crude estimate for the 
maximum field strength of flux tubes before they leave the region towards 
the surface. It is important to note that in the nonlinear dynamical simulations, 
the residence time of flux tubes in the overshoot region 
(Sect.~\ref{ssec:dynamics}) is considerably longer than the linear growth time of 
the instability. The latter timescale, $t_g$, only hints at the e-folding time of linear 
perturbations according to an analytical perturbation analysis. 
As long as the buoyant flux-loss criterion is concerned, field strengths between 
500 to 600 kG would thus be possible to reach, provided that flux-tube explosions 
at $8\Omega_\odot$ can operate for stronger fields than on the Sun. 

\section{Results}
\label{sec:Results}

\subsection{Dynamics of flux tubes near the equator}
\label{ssec:dynamics}

\begin{figure}
    \centering
    \includegraphics[width=.8\linewidth]{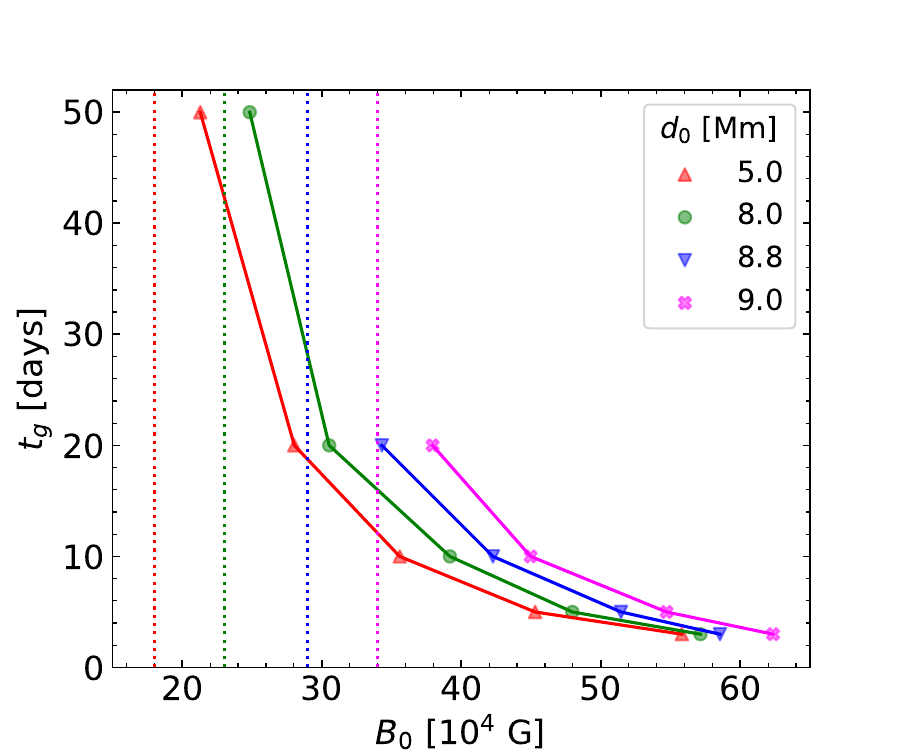}
    \caption{Characteristic growth time $t_g$ for flux 
    tubes in a star with $8\Omega_\odot$, located at 
    depths $d_0$ below the base of the CZ. The vertical 
    dotted lines show, with the respective colours, the 
    onset field strengths of the buoyancy instability at 
    each depth, i.e., with infinite linear growth time.}
    \label{fig:grtime}
\end{figure}

We carried out a series of simulations to compute the trajectory and shape of 
a thin flux tube in a parameter space spanned by the initial depth and field 
strength, as determined by the chosen $t_g$ of the instability. 
The grid of initial parameters chosen are shown 
in Fig.~\ref{fig:grtime}. The curve for $d_0=5$~Mm is the same 
as the red curve in Fig.~\ref{fig:time_rempel}. 
The field strength at a given $t_g$ increases with depth, 
because the ambient stratification becomes more subadiabatic, 
stabilising vertical displacements. As a result, the magnetic 
buoyancy needed to overcome this stabilising effect increases 
with depth. 
The resulting curves are still closely packed together, so 
that the intensification-emergence picture discussed in 
Sect.~\ref{ssec:growthtimes} is only weakly sensitive to 
the initial depth. The deepest case (9 Mm) is only 1000 
km above the radiative zone (the radius of the flux tube 
we consider), where the temperature gradient becomes more or 
less constant with depth.

\begin{figure*}
    \centering
    \includegraphics[width=.45\linewidth]{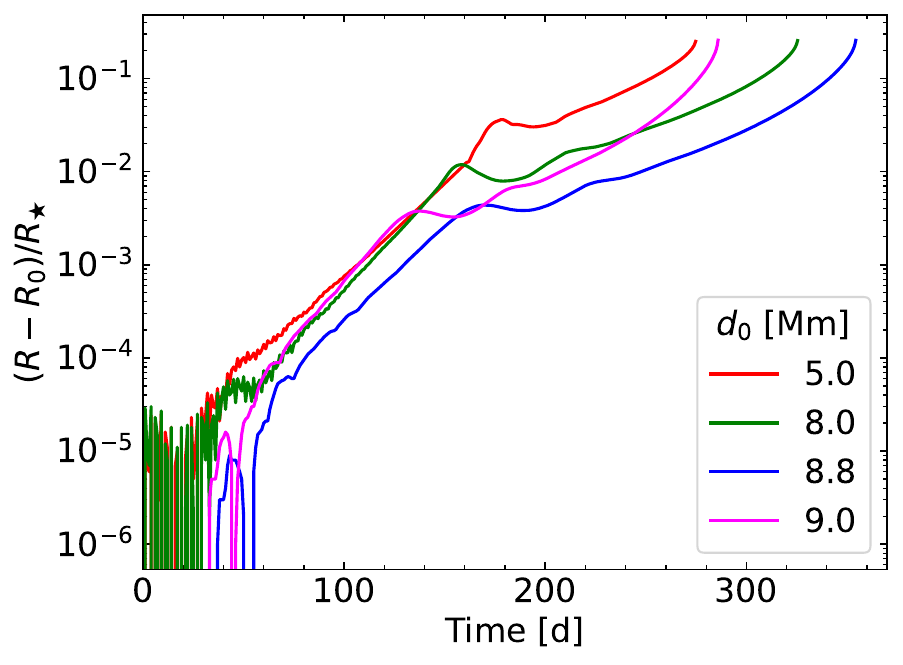}
    \includegraphics[width=.45\linewidth]{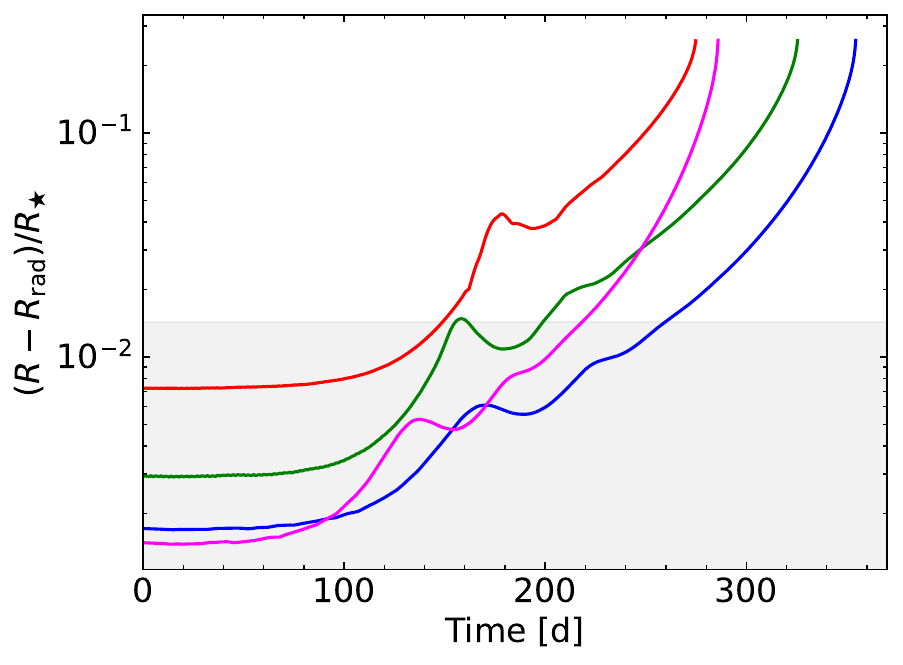}\\
    \includegraphics[width=.45\linewidth]{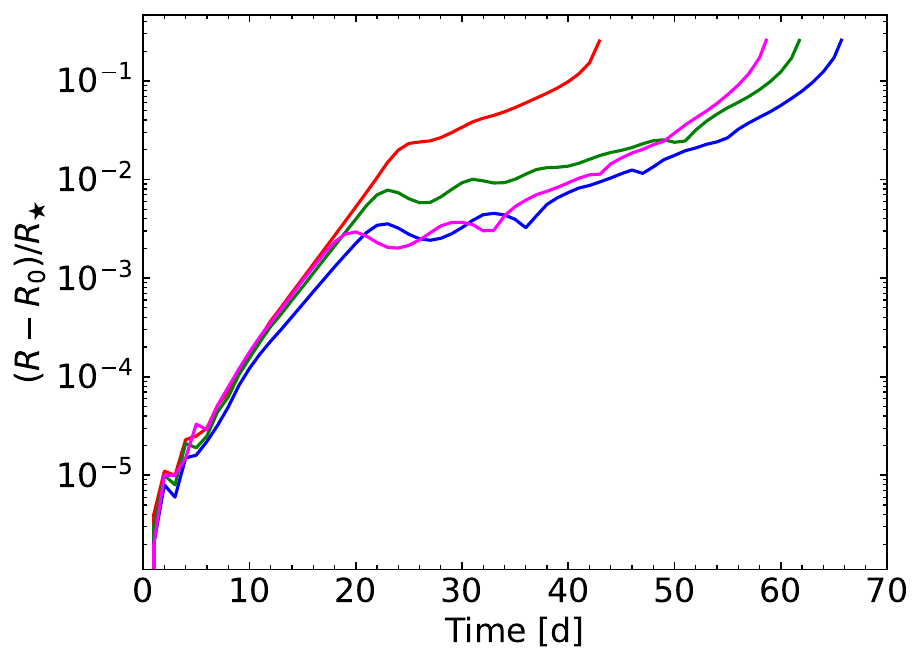}
    \includegraphics[width=.45\linewidth]{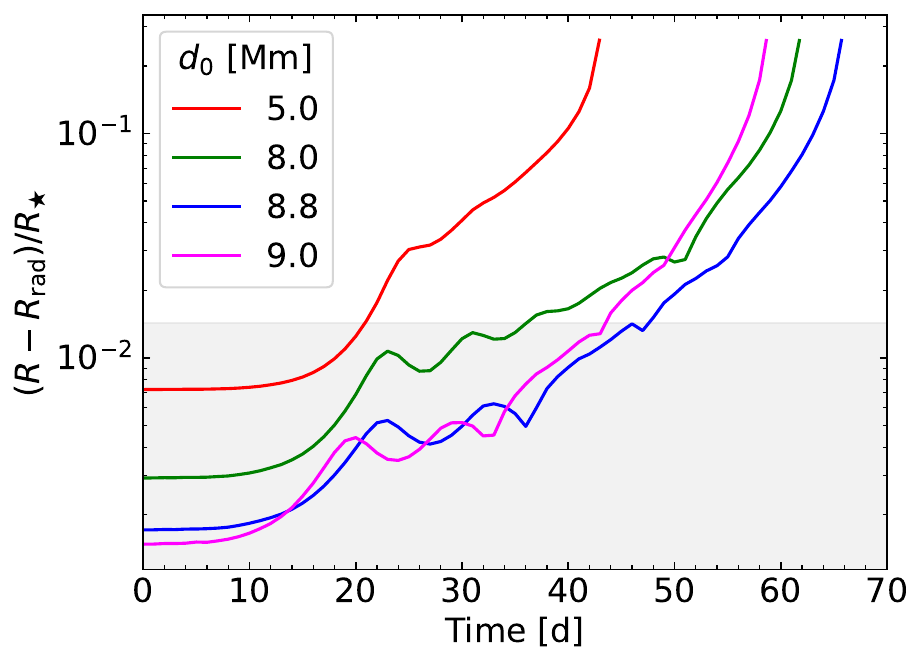}
    \caption{The radial displacement of the tube apex from the 
    initial location $R_0$ (left panels) and relative to the 
    base of the overshoot region (right panels) 
    as a function of time, with different starting depths 
    below the convection zone (colours), for a linear growth 
    rate of 20 days (top panels) and 3 days (bottom panels). 
    The grey horizontal band denotes the overshoot region.}
    \label{fig:rise-t}
\end{figure*}

The radial location of the tube apex as a function of time is shown in 
Fig.~\ref{fig:rise-t}, for a near-critical ($t_g=20$~d) and a highly supercritical 
($t_g=3$~d) cases and various 
initial depths below the CZ-base ($5<d_0<9$~Mm). 
The total rise time of the strongest flux tubes in our sample (i.e., with the 
shortest $t_g$ of 3 days) is between 1-2 months from the initial 
perturbation to the emergence at the surface, leaving 1-2 weeks for the 
nonlinear development of the instability in the overshoot region (shown as a 
horizontal band). Although $t_g\simeq t_f$ (see Sect.~\ref{ssec:growthtimes}), 
the duration of this nonlinear growth phase within the overshoot region is still longer 
than $t_f$ by a factor of three, indicating that there is sufficient time for the 
intensification of the flux tube to a strength which allows for a nearly radial rise 
through the convection zone. For comparison, considering the 
weakest tubes with $t_g=20$~d starting at $d_0=5$~Mm, the time 
it takes for a tube apex to leave the overshoot region is about six months. 

The relation between the linear growth time and the full rise time is 
similar for different $t_g$ values, ie., initial field 
strengths. The deepest-located tubes with $d=9$~Mm emerge earlier than most other cases. 
The reason for this irregular behaviour is 
the nonlinear dynamics (the interplay between various forces acting on each part of the 
tube), which amplify the resonant mode, leading to an early entry to the convection zone 
in some cases. In this particular case for 9 Mm, the buoyancy of the apex becomes large enough to trigger a fast rise before the restoring force acts for a sufficient time. 

\begin{figure*}
    \centering
    \includegraphics[width=\linewidth]{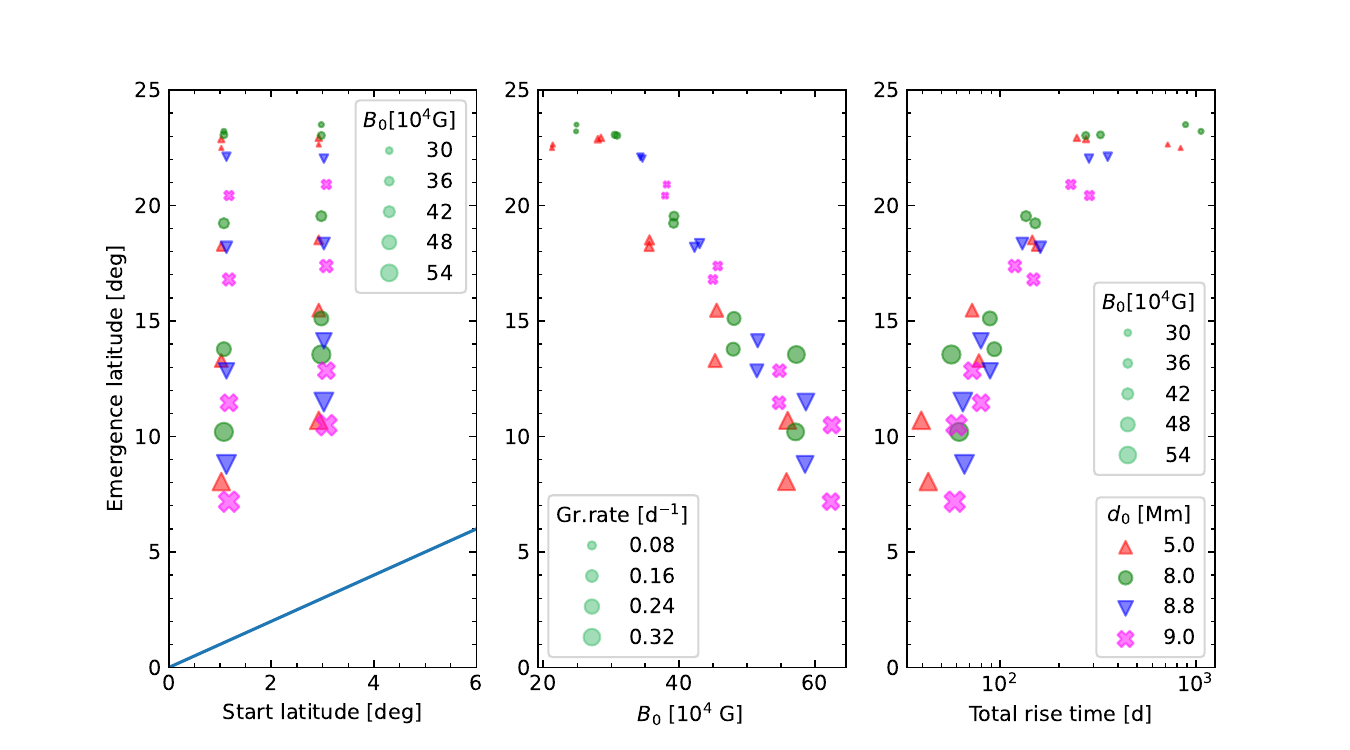}
    \caption{Emergence latitudes of flux tubes as a function of the initial latitude, 
    field strength, and the total rise time (from the beginning of the initial perturbation). On each panel, the marker styles denote the initial 
    depths as in Fig.~\ref{fig:time_rempel}. The symbol size shows the initial field 
    strength on the left and right panels, and the linear growth rate on the middle panel. 
    The blue line shows the radial-rise relation.}
    \label{fig:rise}
\end{figure*}

Figure~\ref{fig:rise} shows the emergence latitude as a function of the start 
latitude, the initial field strength, 
and the total rise time. Shown are simulations with two initial latitudes, 
namely 1 and 3 degrees. The apices of deeper-located and in particular stronger tubes are deflected poleward less strongly, 
owing to a stronger buoyancy and a faster rise. 
The emergence latitudes down to latitudes $\lesssim 10^\circ$ are obtained, 
only in cases where the field strength is above 500 kG, which is about 5 times the critical field strength at the solar rotation rate, and about $2B_{\rm cr}$ for the fast-rotator in consideration. 

The emergence latitude is thus mainly sensitive to the field strength. As a secondary effect, the latitude is in general further decreased by 3-5 degrees with increasing depth at each constant growth rate, for fields above 300 kG, owing essentially to increased field strengths of tubes rooted at greater depth.
The middle panel of Fig.~\ref{fig:rise} depicts how increasing magnetic buoyancy 
consistently brings 
the emergence latitudes from above 20 to below 10 degrees, while increasing the depth adds 
only a few more degrees. The functional dependence 
of $B_0$ on the linear growth rate $t_g^{-1}$ of the instability is also depicted by the 
symbol size. 
The total rise time also indicates how long the Coriolis force 
can act on the rising tube (Fig.~\ref{fig:rise}, right panel). Again, an increased 
magnetic buoyancy ($B_0$ marked by symbol size) leads to a faster rise, reducing the 
relative contribution of the Coriolis acceleration on the resulting rise trajectory. 

Because the sensitivity of emergence latitudes on the initial depth is 
negligible when compared to that in the field strength, we confine the 
analysis in the rest of the study to a single depth ($d_0=5$~Mm), and vary 
$t_g$ only. 

\subsection{Sensitivity to differential rotation}
\label{ssec:dr}

We employed two differential rotation profiles in this study. The first one is the cone-like rotation profile, which approximates the helioseismically inferred solar internal rotation: 
\begin{eqnarray}
    \Omega(r,\lambda)/\Omega_{\star,{\rm eq}} &=& 
    1 + 2\frac{c_3}{\tilde{\omega}} - \left[1+{\rm erf}\left(\frac{r-r_0}{d}\right)\right] \nonumber \\ 
    && \times \frac{1}{\tilde{\omega}}\left(c_1\sin^4\lambda + c_2\sin^2\lambda + c_3\right),
    \label{eq:cone_dr}
\end{eqnarray}
where $\Omega_{\star,{\rm eq}}$ is the rotation rate at the equator, $\tilde{\omega}$ is the stellar equatorial rotation rate normalised to $\Omega_{\odot,{\rm eq}}$, $c_1 = 0.0876$, $c_2 = 0.0535$, and $c_3 = 0.0182$ \citep{isik18}. 
The simulations in Sect.~\ref{ssec:dynamics} were carried out under the 
rotation profile in Eq.~(\ref{eq:cone_dr}).

\begin{figure}
    \centering
    \includegraphics[width=.9\columnwidth]{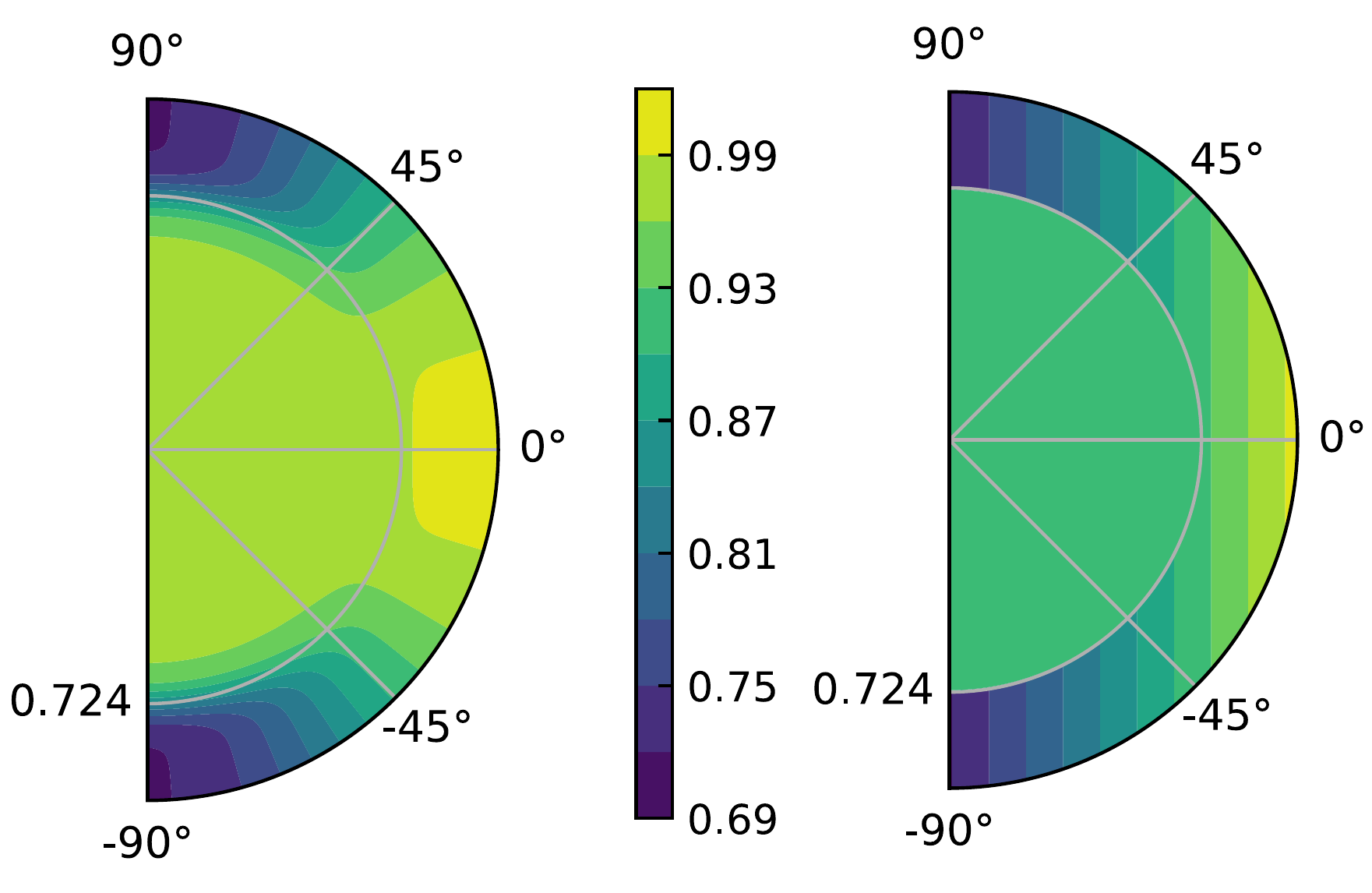}
    \caption{Contours of angular velocity 
    normalised to the maximum at the equator at 
    $r=1$, for solar-like (left panel) and cylindrical (right panel) rotation profiles.}
    \label{fig:dr}
\end{figure}

As an alternative, we assume constant rotation along cylinders, to represent the Taylor-Proudman state of rotation, which is expected in very rapidly rotating stars \citep[e.g.][]{lund+14}: 
\begin{eqnarray}
    \Omega(r,\lambda)/\Omega_{\star,{\rm eq}} =
    \begin{cases}
        \alpha + \Delta\Omega(r\cos\lambda) & {
        \rm if} \quad r > r_{\rm tach} \\
        \alpha+(\Delta\Omega)r_{\rm tach} & {\rm if} \quad r < r_{\rm tach},
    \end{cases}
    \label{eq:cyl_dr}
\end{eqnarray}
where $r_{\rm tach} = 0.724R_\odot$, $\Delta\Omega$ is the pole-equator difference in angular velocity at the surface, set to the same value adopted in Eq.~(\ref{eq:cone_dr}) as the solar reference value. The rotation profile for $8\Omega_\odot$ is expected to be closer to being cylindrical than to conical \citep{ko12, brun22}, so 
we adopt the profile Eq.(\ref{eq:cyl_dr}) for simplicity, as modelling the detailed internal rotation for a given surface rotation rate is beyond the scope of the current study. 
Figure~\ref{fig:dr} shows the rotation contours for both idealised cases. As we show shortly, 
however, the rotation profile does not have a significant effect on 
the emergence latitude. 

From observations, the surface shear is reported to be 
weakly increasing with $\tilde{\omega}$, following a power law $\Delta\Omega\propto \tilde{\omega}^p$, with the exponent ranging from 0.15 
\citep{barnes05} to 1.5 \citep{reiners06}. We adopt $p=0.20$ from the 
literature, which was found for G dwarfs, by interpolating over the 
effective temperature and the rotation rate of a stellar sample 
\citep{balona16}. As an extreme case, we also choose $p=0.66$ following 
\citet{messina+guinan03}, who estimated pole-equator lags from the 
photometrically observed range of rotational periods at a given star. These two 
choices for $p$ corresponds to setting $\Delta\Omega$ in Eq.~(\ref{eq:cyl_dr}) 
to about 1.5 and 4 times the solar value. 

\begin{figure}
    \centering
    \includegraphics[width=\columnwidth]{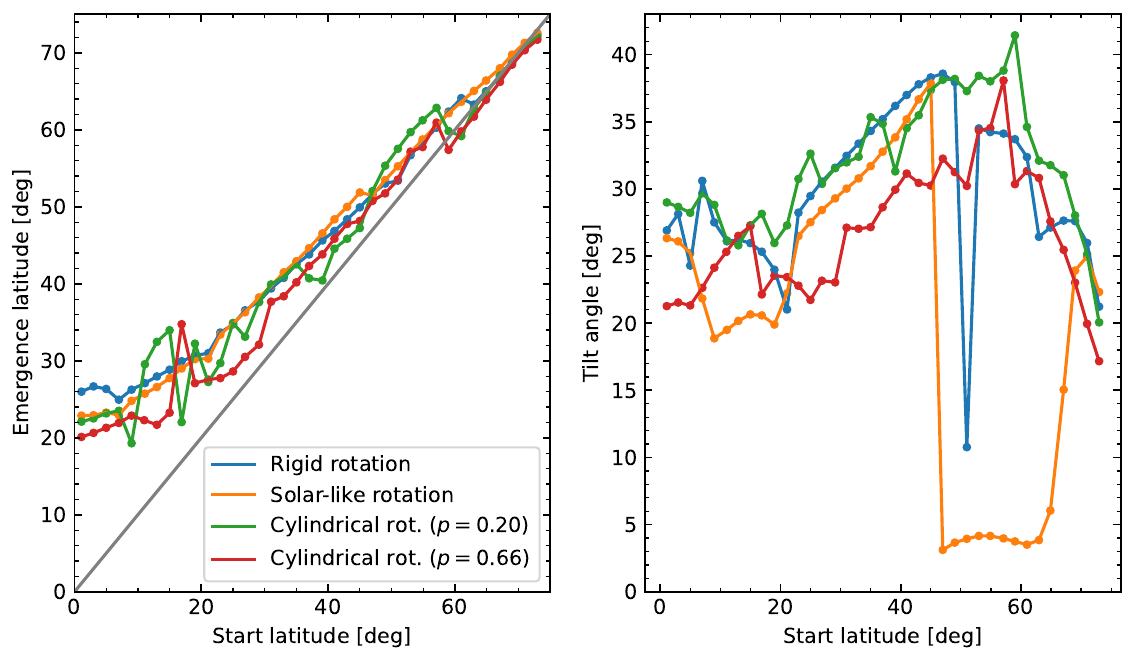}
    \caption{Emergence latitudes (left panel) and tilt angles (right panel) 
    as a function of the initial latitude, at an initial depth of $d_0=5$~Mm below 
    the base of the convection zone and $t_g=20$~d. Note the relatively strong poleward 
    deflection for low initial latitudes.} 
    \label{fig:DRcomparison}
\end{figure}

A comparison of initial and emergence latitudes and the tilt 
angle distributions of flux tubes are given in Fig.~\ref{fig:DRcomparison} for rigid-body, 
solar-like, and cylindrical rotation laws (the latter for $p=0.20$ and 0.66), with 
the initial depth $d_0=5$~Mm and linear growth time $t_g=20$~d. 
The results are not very sensitive 
to the differential rotation profile except for somewhat smaller poleward deflections over a wide range of initial latitudes in the case $p=0.66$. This justifies our use of solar-like rotation for the purposes of the 
earlier studies \citep{isik11,isik18}. The large tilt angle jumps 
in the solar-like rotation case were driven by the occurrence of two 
adjacent loops with different curvatures \citep[][Fig.~5]{isik18}. Such 
configurations do not form under the cylindrical rotation profile 
and the resulting tilt angle dependencies are relatively smooth.

\begin{figure}
    \centering
    \includegraphics[width=\columnwidth]{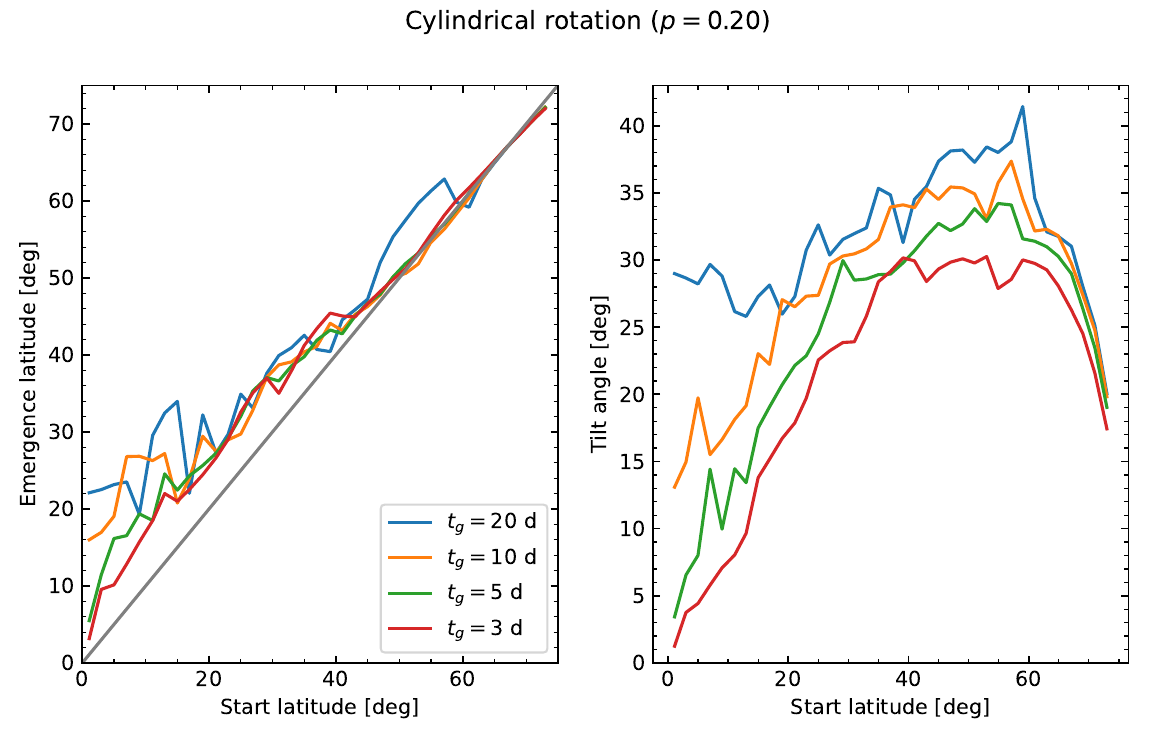} \\
    \includegraphics[width=\columnwidth]{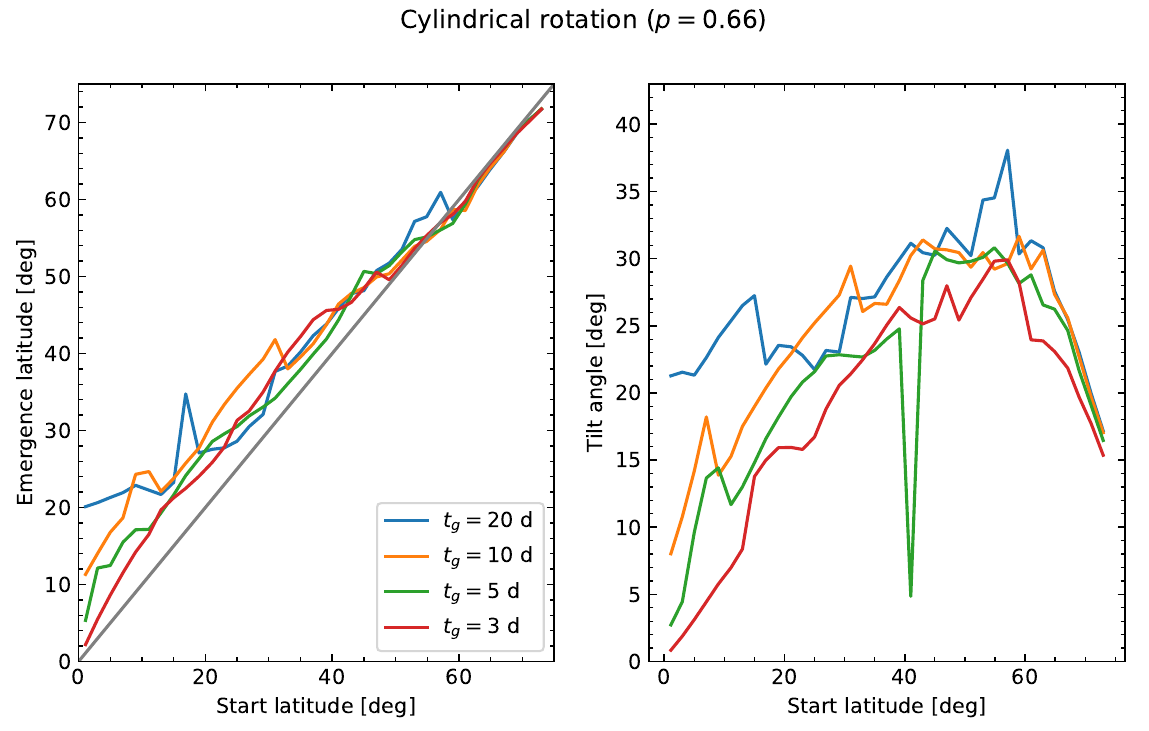}
    \caption{Similar to Fig.~\ref{fig:DRcomparison}, for different 
    initial magnetic field strengths corresponding to the indicated 
    linear growth times, for $p=0.20$ (top panel) and $p=0.66$ (bottom 
    panel).}
    \label{fig:cylDRemergence}
\end{figure}

The results of flux-tube rise simulations for various initial field strengths and the two shear amplitudes (for $p=0.20$ and $p=0.66$) are presented in 
Fig.~\ref{fig:cylDRemergence}. Poleward deflection of flux tubes is 
much suppressed by buoyancy in the case of short growth times 
(greater initial field strengths), leading to almost radial rise for 
the shortest growth times. 
The emergence latitude is rather insensitive to the change in the 
differential rotation amplitude (from 1.5 to 4 times the solar value), 
whereas the tilt angles show an overall suppression by a few degrees. 

\subsection{Latitudinal distributions of emergence}
\label{ssec:patterns}

Next, we investigate the main implications of highly supercritical flux 
tubes in shaping the latitudinal distribution of starspot emergence. To do so, 
we assumed three probability distribution functions (PDFs) for the start latitudes of tubes 
at the base, and interpolate their nonlinear mapping to the surface, using 
the simulation results for $\tilde{\omega}_\star=8$ in Sect.~\ref{ssec:dr} 
under the cylindrical rotation profile, Eq.~(\ref{eq:cyl_dr}), with the amplitude scaling exponent $p=0.20$. The interpolation table spans initial latitudes from 
$1^\circ$ to $73^\circ$ and has an angular resolution of $2^\circ$. The upper 
limit of latitude is set by the linear instability boundary on the $(B_0,\lambda_0)$ 
plane.

\begin{figure}
    \centering
    \includegraphics[width=\columnwidth]{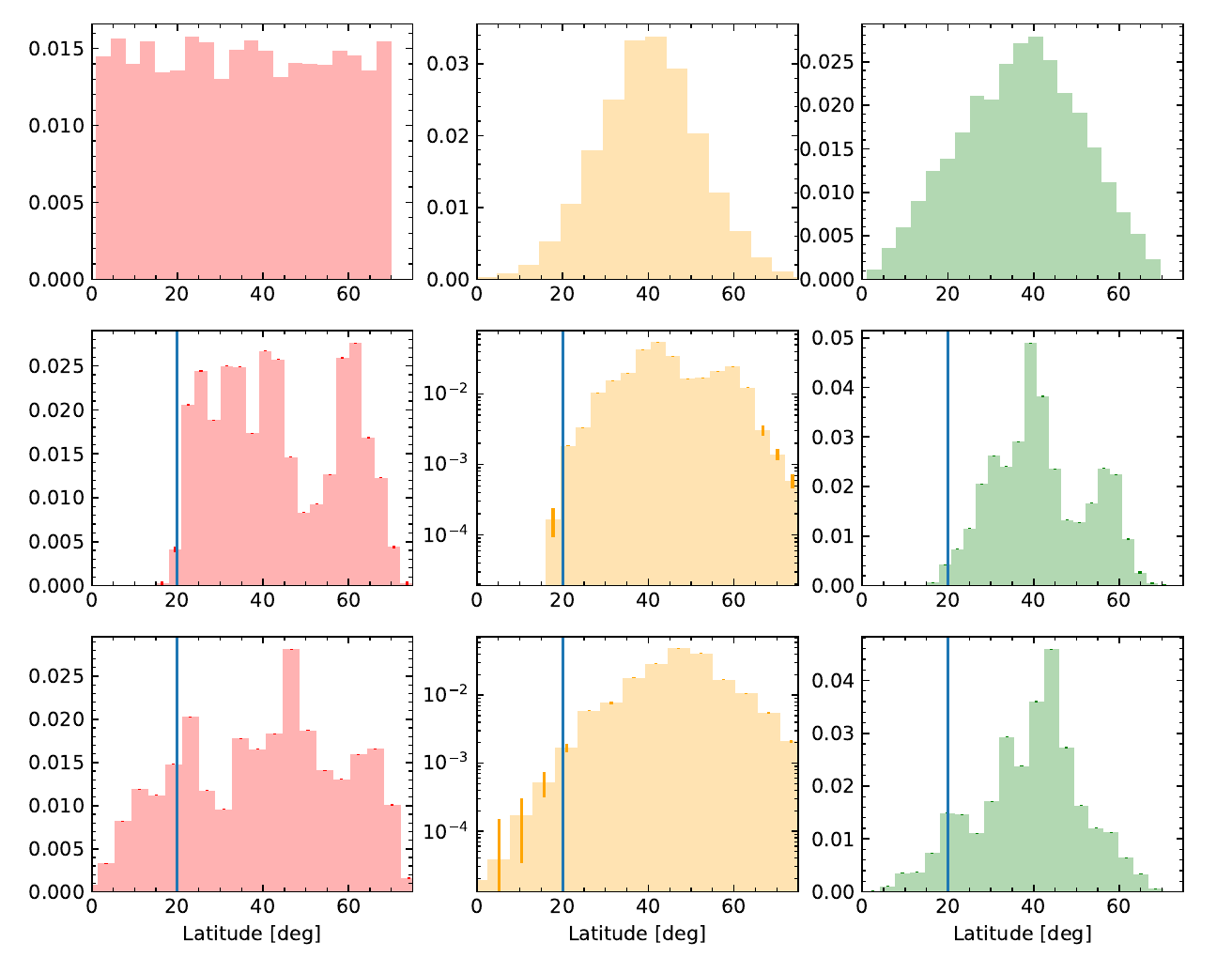}
    \caption{Normalised histograms for start latitudes at a depth of 5 Mm below the base of the convection zone (top panel; with uniform, normal, and triangular distributions), compared to emergence latitudes for near-critical ($t_d=20$~d) and super-critical ($t_d=3$~d) initial field strengths. The vertical lines are drawn for evaluating low-latitude emergence.}
    \label{fig:histos}
\end{figure}

We produce $10^4$ random realisations, each following one of the uniform (between $1^\circ$ and $70^\circ$), normal ($\mu=40^\circ$, $\sigma=12^\circ$), and triangular (peak at $40^\circ$ in ($1^\circ, 70^\circ$)) PDFs for the start latitudes in the overshoot region, as shown on the top row of Fig.~\ref{fig:histos}. We then interpolate the mappings in Fig.~\ref{fig:cylDRemergence} for $p=0.20$, to obtain the corresponding emergence latitude distributions, with an additive Gaussian noise in latitude, with $\sigma=2^\circ$, to simulate possible effect of convective buffetting during the rise. When near-critical field strengths corresponding to $t_g=20$~d are chosen, the emergence is confined above $\sim 20^\circ$ latitude (Fig.~\ref{fig:histos}, middle row). For higher field strengths, the minimum latitude decreases, guided by the emergence latitudes in Fig.~\ref{fig:cylDRemergence}. For the shortest linear growth time that we consider (3 d; bottom panel), significant flux emerges between the equator and $\sim 20^\circ$ latitude. When the initial distributions are Gaussian or triangular, the fraction of emerging flux near the equator is less than for the uniform base distribution. These results hint at the possibility to estimate base distributions of toroidal flux using observed surface distributions, within a probabilistic inference framework that would also involve the surface flux transport layer of the FEAT model \citep{isik18}. 

To test whether the substructure in the surface latitudinal distributions are not due to noise, we derived standard errors ($\sigma/\sqrt{N}$) from 20 independent realisations for each latitude bin (summing to $2\times 10^5$ realisations per histogram). The resulting error bars are too small to be 
visible in most of the histogram bins. 
The multiple peaks in the surface distributions result from the 
non-uniform mapping between the base and the surface, making an inherent 
signature of the dynamics of flux rise on the surface. Such non-uniformities 
are also visible in some (azimuthally averaged) latitude distributions of active-star Doppler images similar to Fig.~\ref{fig:EKDraprofile}. However, the latter distributions are possibly contaminated by the low resolving power in latitude and possibly by the continuous redistribution of emerging flux on the stellar surface by horizontal transport processes.

\section{Discussion}
\label{sec:discuss}

In this study, we calculated the magnetic flux tube configurations which can give rise to low-latitude emergence of 
magnetic flux in a solar-like convection zone rotating eight times faster than the Sun 
($P_{\rm rot, eq}=3.125$ days). 
To shift the lowest emergence latitude of flux tubes 
originating from the base of the convection zone from above $\sim 20^\circ$ to about $7^\circ$ for solar-like conical shear, or $3^\circ$ for cylindrical shear, our computations require field strengths 
that are stronger than the near-critical ($B\sim B_{\rm cr}$)
levels ($t_g=20$~d) by a factor of about 2. The same supercriticality 
factor is about 1.6 near the base of the overshoot region ($d_0=9$~Mm). 

We studied two main effects that are likely to determine the 
initial field strength of flux tubes near the base of the convection zone. 
As the amplifying source, we considered explosions of flux tubes carrying 
significantly low flux in the superadiabatic part of the convection zone 
\citep[see also][for the impact of convective flows on thin flux tubes]{weber11}. 
For this effect to be functional up to field strengths compatible with 
low-latitude emergence, we found that the bulk of the unstably stratified 
part of the convection zone must be more superadiabatic than in solar models, 
by a factor of about 
three, so that explosions of $10^5$-G tubes (with presumably lower fluxes 
than non-exploding ones) would be possible within the CZ. This requirement is 
consistent with the global MHD simulations of rotating convection, 
indicating a reduction in the efficacy of convection with increasing Coriolis number. 
In this way, low-latitude-emerging tubes with $B_0=500$~kG can be formed. 
The second important effect is the magnetic buoyancy instability, 
whose growth rate increases faster with the tube strength than the 
amplification rate (see Sect.~\ref{ssec:growthtimes}, Fig.~\ref{fig:time_rempel}). 
Because the unstable flux tubes require considerably longer to break out 
of the overshoot layer than the linear growth-time constraint 
(Sect.~\ref{ssec:dynamics}), we expect the buoyancy constraint not to strongly 
impede the amplification process in the nonlinear regime, for $B_0\lesssim 500$~kG. 

Doppler images of rapidly rotating G-K stars often show low-latitude spot concentrations, but generally only a few at a given time, as opposed to much higher 
spot coverage at mid- to high latitudes. 
In the context of our results, we can 
interpret the observed latitudinal distribution of starspots as follows: 
starspot-producing flux tubes can form over a range 
of field strengths at various depths in the overshoot layer, with a given probability density of formation at a given field strength. 
Formation of exceptionally strong flux tubes have more radial rise trajectories than weaker ones. If the flux tube in the overshoot layer was ejected from near the equator, then this can lead to a low-latitude active region. 
The flux-rise mapping of the most buoyant tubes in our sample shows that even for a uniform base distribution (Fig.~\ref{fig:histos}, bottom panel), the surface distribution declines towards the equator, consistent with Doppler imaging results. 

The dynamo generation of toroidal flux and the surface patterns driven by flux emergence and surface transport are all likely to contribute to the 
observed patterns of surface activity \citep{isik11}.
The surface manifestations of stellar activity will be partly determined by the distribution of toroidal flux in the convection zone, from which starspot-producing flux loops would originate. Knowing the toroidal flux distribution requires better understanding of dynamos at various rotation rates \citep[see, e.g.,][]{Finley+24}. 
By taking toroidal flux distributions in the convection zone and accounting 
for low-latitude emergence of magnetic flux on stars 
rotating up to 10 times the solar rate, the distribution and evolution of 
starspots can be modelled, using surface flux transport models. 
The resulting surface maps can be used in forward-modelling observational 
diagnostics, in comparison with observations 
\citep[e.g.,][]{sowmya+22,nemec23}.

In a recent study, \citet{Zhang+Jiang24} introduced a rotation-dependent mean latitude for the surface source of a flux-transport dynamo model, to find shorter activity cycles and a mode change of the global field from dipole to quadrupole as the rotation period decreases. Motivated by our results, we note that low-latitude emergence of active regions near the equator can have a constructive effect on the global dipole even for fast rotators, if the Coriolis-induced tilt angles of emerging bipoles or the random impact of convective motions are large enough \citep[][and references therein]{weber23}. In the current study, we find a tilt angle of $3^\circ$ for the minimum emergence latitude reached ($7^\circ$) in the $d_0=9$~Mm case, and $1-5^\circ$ for the cylindrical-rotation cases and $t_g$ of 3-5 days. 

We have assumed that the magnetic flux that forms stellar active regions has its source near the base 
of the convection zone, where the stratification is sufficiently stable 
for super-equipartition (and super-critical) flux 
tubes to form. However, the thin flux tube model has several limitations. Relevant to the present context, (a) the flux emergence dynamics from the subsurface into the photosphere is out of its scope; (b) the sizes of solar active regions are smaller than what is predicted by the model, where the unstable loop has a considerably longer azimuthal extent (sunspot groups can also be fragments of larger-scale loops that form in the convection zone). Convective and rotational effects within that skin depth near the surface can also have substantial effects in the formation of solar and stellar active regions \citep[][and the references therein]{weber23}. Although no studies have so far ruled out the possibility for sunspot groups to originate from the base of the convection zone, an alternative flux emergence mechanism can be led by a distributed dynamo \citep[][and the references therein]{kaepyla23}, 
where flux tubes can form in the midst of the convection zone, e.g., out of shear-generated wreaths and they can rise to the surface, subject to buoyancy and convective flows 
\citep{nelson11,nelson14,weber23}. 
In such convective dynamo action, flux tubes starting from low latitudes not very far below the surface
could emerge at low latitudes more easily, owing to the relative proximity of the source region to the surface. 

\section{Conclusion}
\label{sec:conc}

We studied the storage, intensification, and rise properties of magnetic flux in a solar-type
star rotating with a period of about 3 days, within the thin flux tube approximation. We find,
\begin{itemize}
\item The initial field strength required for the low-latitude emergence is about 500~kG. Flux tubes can be amplified to such strengths in a series of failed emergences (explosions) of weaker flux tubes, provided that the temperature gradient in the upper convection zone increases with the rotation rotation rate. 
\item Field strengths of order 500~kG can be established inside flux tubes before they would be lost from the overshoot region. 
\item The suppression of poleward deflection is more sensitive to the field strength than to the depth at which a flux tube is located in the overshoot layer.
\item The results have only a weak dependency on the detailed shape and amplitude of internal 
differential rotation. 
\item Various initial latitude distributions of flux tubes end up with similar emergence distributions, peaking around some latitudes and showing a tail towards the equator, similar to Doppler imaging observations. 
\end{itemize}

Our results will have implications 
for the forward modeling of observational diagnostics of active stars \citep[see][Sect.~3]{isik23} 
and attempts to constrain stellar dynamo models \citep[see][Sect.~11]{charbonneau+sokoloff23}. One consequence of taking into account low-latitude emergence in 
active stars would be a better estimation of the contribution of faculae to 
brightness variations in the rotational and activity-cycle time scales 
\citep{nemec22}. Another implication would be in the forward modelling of 
chromospheric activity indices for active stars, which we plan to 
cover in an upcoming study.

\begin{acknowledgements}
     {\bf Acknowledgments.} We thank the anonymous reviewer whose critical comments helped us to 
    improve the manuscript considerably. We acknowledge Hakan V. \c{S}enavc{\i} for providing the Doppler-imaging map of EK Dra in Fig.~\ref{fig:EKDraprofile}. 
\end{acknowledgements}

\bibliography{deepOS}{}
\bibliographystyle{aasjournal}



\end{document}